\documentstyle[preprint,aps]{revtex}

\tightenlines

\begin{document}

\title{Study of Vortex Avalanches using the Bassler-Paczuski Model} 
\author{G. Mohler and D. Stroud}
\address{
Department of Physics,
The Ohio State University, Columbus, Ohio 43210}

\date{\today}

\maketitle

\abstract{We have carried out a model calculation of the flux noise produced
by vortex avalanches in a Type-II superconductor, using a simple kinetic 
model proposed by Bassler and Paczuski.  Over a broad range of 
frequencies, we find
that the flux noise $S_{\Phi}(\omega)$ has a power-law dependence on frequency,
$S_{\Phi}(\omega) \sim \omega^{-s}$, with $s \sim 1.4$  in reasonable
agreement with experiment. In addition, for small lattices, the
calculated $S_{\Phi}(\omega)$ has
a high-frequency knee, which is seen in some experiments, and which is due
to the finite lattice size.  Deviations
between calculation and experiment are attributed mostly to uncertainties in the
measured critical current densities and pinning strengths of the experimental
samples.

\section{Introduction}

Bassler and Paczuski\cite{BP} have recently proposed a model which simulates
the development of the Bean critical state in a type-II superconductor, as well
as of other types of so-called self-organized critical behavior~\cite{bak}.  
In such a system, vortices are injected into one edge of an initially empty
sample, and are free to fall off the opposite edge only.  As vortices
are driven into the system from the ``loading edge,'' they push 
already-present vortices further into the sample.   These existing vortices
then pile up in such a way that the vortex density gradient approaches a 
critical value which is determined by system
parameters such as pinning strength and density.  Once the gradient reaches
this critical value, the system is said to have achieved a ``self-organized
critical state.''  

Of particular interest in this context, is the development of ``avalanches''
from the Bean critical state.  When the flux density gradient 
exceeds its critical value,
the injection of even a single excess vortex into the system
can start a chain reaction of vortex motion, known as an avalanche, 
which may have a very large scale in both space and time and which 
causes the gradient to relax back to its critical value.  Vortex avalanches 
are typically characterized by their duration, their linear extent 
in the direction of average vortex motion, and the number of vortices forced 
from their original positions (``topplings'') by the avalanche.  These
and other avalanche characteristics are expected to obey various
scaling laws~\cite{bak2}.

In earlier numerical models, the generation of vortex avalanches was 
typically studied numerically at a ``microscopic'' scale, i.\ e.,
at a length scale where individual vortex displacements were
calculated from certain postulated force laws\cite{Nori}.  
In practice, this type of model imposes severe constraints on the size
of system which can be studied numerically; such constraints 
in turn make it difficult to study the critical behavior of vortex avalanches.
But since an avalanching system is expected eventually to achieve
a self-organized critical state,
the large-scale behavior of the system should be describable without tracking
the motion of individual vortices.  The BP model takes
advantage of this expectation by focusing attention only on the large-length
scale behavior of this system, which can be modeled using only a coarse-grained
lattice.  This choice makes feasible the simulation of
large systems.

\section{The Bassler-Paczuski Model: Definition and Method of Calculation}

In the Bassler-Paczuski (BP) model, one considers a distribution of
vortices on a two-dimensional simple hexagonal lattice (see
Fig.~\ref{fig:lattice}).  The scale of the lattice is 
assumed to be such that the vortices can be treated as point-like objects
which exist entirely on the lattice sites.  Each lattice site is capable
of containing multiple vortices, and the number of vortices per
lattice site is a kind of coarse-grained vortex density\cite{note1}.  
Each lattice site is
also described by a pinning potential, which is chosen at random from a
suitable distribution.  Periodic boundary 
conditions are imposed at the top and bottom of the lattice.  Vortices
are injected into the left-hand edge, and removed from the 
right-hand edge, as described in more detail below.
As more and more vortices are injected into the lattice, 
the repulsion between vortices eventually
overcomes the attraction exerted on the vortices by the pinning sites.  
As a result, the vortices slowly migrate from left to right
across the lattice.  When they reach the right-hand edge, in this model, 
they are assumed to ``fall off'' the lattice and are removed from 
the system.  Vortices are forbidden to exit from the left side of the system,
or to re-enter the lattice from the right-hand side once they fall off.  

An important parameter in the BP model is the rate of vortex injection. 
In the ``slow-driving limit''\cite{BP}, a vortex is injected into the
lattice only after an avalanche has ceased.  This slow-driving
limit is similar to that studied in models which treat the microscopic
(i. e., the individual vortex) degrees of freedom together with
explicit force equations for each vortex.  For example, 
Nori {\it et al}\cite{Nori}, who have studied such a model, introduce a new
vortex into the lattice only when the lattice-averaged vortex velocity 
has fallen below a certain threshold\cite{Nori}.  In the present paper,  
we inject vortices either slowly or quickly, so that we may make predictions
for either a hypothetical slow-driving regime or the high magnetic
field ramping rates studied in most experiments\cite{exp}.

In the BP model, the force acting on a vortex at site $x$
in the direction of site $y$ is taken to be
\begin{equation}
F_{x \rightarrow y} = -V(x) + V(y) + [m(x) - m(y) -1] 
		+ r[m(x1) + m(x2) - m(y1) - m(y2)].
\end{equation}
Here $x1$ and $x2$ are the two sites (other than site $y$) which neighbor $x$ 
on the hexagonal lattice;
$y1$ and $y2$ are the two sites other than $x$ which are nearest
neighbors to $y$; $V(z)$ is the strength of the pinning potential on site $z$; 
and $m(z)$ is the vortex population at site $z$ (taken to be a positive
integer).  $r$ is the strength of the repulsion between vortices on
neighboring sites, in units such that
the on-site repulsion strength is normalized to one.

The BP model is defined by three parameters, denoted $r$, $p$, and $q$.  
$r$ is the nearest-neighbor repulsion parameter mentioned above, 
which is assumed to be smaller than unity;  
$p$ is the pinning strength; and $q$ is the probability that a
particular site contains a nonzero pinning potential.  
Thus, any particular site has
a pinning strength of $p$ or $0$ with probability $q$ and $1-q$ respectively.  
The distribution of pinning sites is determined at random once 
for the entire lattice, at the start of the simulation, and is 
held constant thereafter.

A vortex move is carried out as follows.  First, the force on a vortex is
calculated for each of the three possible directions of vortex motion.
If exactly one of the three forces is greater than zero (i. e., acts towards
a nearest neighbor site), then one vortex is moved by one lattice
spacing in the direction of that force. If more than one force is greater 
than zero, the direction
of motion is chosen randomly from the set of positive forces.  This calculation
is carried out for each site containing at least one vortex, and all the
vortex populations are updated in parallel - that is, each site is examined
once in a given time step, and at most one vortex is moved from
a particular site during a time step.

To carry out the calculations, we constructed
a simple hexagonal lattice using the unit cell shown in 
Fig.~\ref{fig:lattice}, with cell length $a_x$ and
cell height $a_y = \sqrt{3}a_x$.
The total lattice size was chosen so as to have dimensions 
$L_xa_x \times L_ya_y$, as shown in the Figure, with $L_x = 4L_y$, and
$L_y$ integer.  For this choice of dimensions, and for $L_x > 32$,
the rectangular sample was found to be wide enough to prevent individual
avalanche events from overlapping one another in space.
For a given lattice size, we carried out simulations of avalanche formation and
evolution for a range of pinning strengths and vortex injection rates.

\section{Results}

\subsection{Numerical Results Obtained from BP Model}

All of the runs were carried out using $2^{20} \sim 10^6$ time steps.
For most choices of parameters and vortex injection rates, it was found
that the vortex population quickly reached a 
steady state of the form predicted by the Bean model~\cite{bean1,bean2}.  That
is, the magnetic induction gradient $dB/dx$
from left to right (i. e., in the
direction of vortex injection) rapidly approached a constant value (see Table).
According to the Bean model, \cite{beant} this constant value is related
to the critical current density $J_c$ of the superconductor
(see below).

Our goal in this work is to calculate the flux noise 
produced by vortices falling off the right edge of the sample.  This goal
is similar to that of Jensen, who has also used a coarse-grained
model to study the power spectrum of a self-organized critical
state\cite{Jensen}.  Jensen's model, however, differs from the BP model in
several respects.  For example, the Jensen model forbids
occupancy of a site by more than one vortex, and it also allows vortices to 
fall off both the right and left edges of the system.
Thus, we expect the two models to give different predictions for noise
spectra.

To calculate the flux noise in the present model, 
we used the following approach.  In experiment, the flux
noise is measured by placing a detector coil in the center of a hollow sample 
ring~\cite{exp}, as shown in Fig.~\ref{fig:sample}.  
According to Faraday's Law, the voltage generated in the coil 
is proportional to the rate of change of vortex population within the coil area.
To display this, we plot in the Figures the 
Fourier transform
\begin{equation}
S_{\Phi}(f) = Lim_{T \rightarrow \infty}|\int_{0}^{T}N^{\prime}(t)
\exp(-i 2 \pi f t)dt|^2,
\end{equation} 
where $N^{\prime}(t)$ represents the rate of change of vortex population 
in the region to the right of the sample at time $t$.  (Our transverse periodic
boundary conditions correspond to a ring geometry.)  In our calculations,
$N^{\prime}(t)$ is taken as the number of vortices which fall off the
right-hand edge of the sample in a single time step at time $t$.  
We evaluate the Fourier transform using the integral given above, then smooth
the resulting power spectrum using a Savitzky-Golay filter as described
in Ref.~\cite{numrec}.   We used first order smoothing over 
twenty points on either side of the data point.

Our results are shown in Figs.~\ref{fig:pow} and~\ref{fig:knee}.  
In general, for all the spectra
shown, the power spectrum exhibits a power-law dependence on frequency,
over a portion of the frequency range, i.\ e.,
$S_{\Phi}(f) \sim f^{-s}$, resembling 
that seen in the experimental results of Field {\it et al}~\cite{exp}.  
In Fig.~\ref{fig:pow},
$s \sim 1.0$ for spectra $a$, $b$, and $c$.  $s \sim 1.4$ for $d$.
In general, $s$ is found to increase from 1.0 to 1.4
when the injection rate is reduced 
below one vortex per time step across the entire left-hand edge.
However $s$ never achieves the experimentally observed value of $\sim 1.5$,
as seen, for example, in the results of  ~\cite{exp}.

For the relatively small lattices, $L \leq 32$, we often
see a characteristic feature in the power spectrum which is 
missing at larger sizes.  Namely, for an injection rate of 
one vortex per time step, where the vortex can enter anywhere
along the left-hand edge,
the power spectrum exhibits a plateau; see Fig.~\ref{fig:knee}($a$).
This plateau disappears as the injection rate
is decreased.  For example, at an injection
rate of 1 vortex per 100 time steps [cf.\ Fig.~\ref{fig:knee}($b$)], 
no plateau is apparent.  The plateau is due to the 
emergence of a new length scale in 
the lattice for small lattices and large injection rates, namely, the
smallest linear dimension of the lattice itself.  Under such
conditions, the lattice size limits the avalanche dimensions.  
This new length scale in turn produces the high-frequency ``knee'' in the 
noise spectrum.  Specifically, at these frequencies, lattice-wide
avalanches dominate, drowning out the high-frequency noise produced
by sporadic independent avalanches.
As in the other power spectra
generated by the model, the calculated slope 
$b \equiv - d ln(S_v(\omega))/d ln(\omega)$
for frequencies $\omega$ above the knee
of Fig.~\ref{fig:knee} is smaller than the experimental value b$_{exp}$;
our calculated value in this regime is $b\sim -1$ compared to the
experimental value of $b_{exp}\sim -2$ ~\cite{exp}.

A vortex injection rate greater than 1/1 (i.\ e., greater than one vortex
per lattice site per time step) was also attempted, so that multiple 
vortices were randomly injected into the lattice simultaneously.  The goal of
this test was to check if a ``knee'' could be created at larger length scales
at sufficiently high injection rates.   But this test resulted in a power
spectrum with nearly white noise, i.\ e., $b \sim 0$.
It is possible that, with the rather limited lattice size of
$L=32$, the size of avalanches is inevitably restricted purely by size
limitations, whereas in the larger lattices,
the vortices exhibit a different avalanche pattern, unconstrained
by lattice size.

\subsection{Connection to Experiment}

In order to compare our model results with the experiments of 
Field {\it et al}, we need to estimate the lattice constant and time step of 
the simulation.  To obtain the lattice constant, 
we look at a cross-section of the lattice, as seen in
Fig.~\ref{fig:bean}.  Then integrating Ampere's Law,
\begin{equation}
	{\bf \nabla} \times {\bf B} = \frac{4\pi}{c} {\bf J}
\end{equation}
along the path in Fig.~\ref{fig:bean}, and assuming a constant current
density equal to the critical current density, $J = J_c$ 
(as expected in the Bean critical state), we find
\begin{equation}
	B_z(x) = B_z(0) - \left(\frac{4\pi}{c} J_c\right)x.
\end{equation}
If we multiply this expression by the area of a narrow
vertical strip of the lattice, $\frac{L}{4}a_ya_x$, we obtain
the flux through the vertical strip centered at $x$ and of width $a_x$ as
\begin{equation}
	\Phi(x) = \Phi(0) - \left(\frac{{\pi}J_cL}{c}a_ya_x\right)x.
	\label{eq:phi}
\end{equation}
To proceed further, we write $x = n_xa_x$, where
$a_x$ is the distance between two opposite sides of the hexagonal cell
(see Fig.~\ref{fig:lattice}), and we denote by $m(x)$
the total flux through a column parallel to $y$, of width $a_x$, 
and centered at $n_xa_x$. 
Then it follows from eq.~\ref{eq:phi} that
\begin{equation}
	m(n_{x}) = m(0) - \left(\frac{{\pi}J_cL}{\Phi_0c}{a_x}^2a_y\right)n_x
\end{equation}
Finally, if the cells are equilateral hexagons of side a, 
then $a_x = a\sqrt{3}$ and $a_y = 3a$, and hence
\begin{equation}
        m(n_{x}) = m(0) - \left(\frac{9{\pi}J_cLa^3}{\Phi_0c}\right)n_x
\end{equation}

We can use this equation to interpret our numerical results for the
vortex densities and flux noise spectrum.  We also need an estimate of $J_c$,
the critical current density for the sample studied experimentally.
In this case, the sample is the composite material $Nb_{0.47}Ti_{0.53}$
(typically in the form of a solid solution with precipitates which
constitute the pinning centers).
Experiments suggest a value in the range of 
$ 3 \times 10^{15} $ statamperes/cm$^2$
~\cite{nbti}, from which we can estimate $a$ from
\begin{equation}
	a  = \left(\frac{m(0)\Phi_0c}{9{\pi}L^2J_c}\right)^{1/3}
\end{equation}
Some representative values of $a$, as calculated from our simulations, are
shown in Table I.

Once $a$ is determined, the simulation time constant,
$\tau_0$, can be estimated by equating the Lorentz force per unit length on
a vortex to the pinning force per unit length, i.\ e.\
\begin{equation}
	f_{pin} = f_{L}
\end{equation}
We assume that $f_{pin}$ can be interpreted as a vortex drag force proportional
to the average vortex velocity $v$, i.\ e., $f_{pin}  = {\eta}v$, 
where $\eta$ is an effective friction coefficient.
Then writing out the Lorentz force on a unit length of a single vortex 
explicitly gives:
\begin{equation}
	{\eta}v = \frac{J\Phi_0}{c}
\end{equation}
Now a vortex can move only one lattice constant per time step, so that
$v \sim a/\tau_0$.  Furthermore, 
$\eta = \frac{B\Phi_0}{{\rho}c^2}$, where $B$ is the local magnetic induction,
$\rho$ is the flux flow resistivity, and $c$ is the speed of light.  
Combining these relations gives
\begin{equation}
	\tau_0 = \frac{Ba}{{\rho}cJ_c}.
\end{equation}
If we assume a magnetic induction of 5 kG, a lattice constant of
$3.22\times 10^{-5}$~cm as suggested by the estimates in the Table, 
a critical current of $3\times 10^{15}$ statamps/cm$^2$ as is typical for the
experiments of Field {\it et al}, and 
a flux flow resistivity of $1.11\times 10^{-23}$ ~sec, as suggested by
experiments on NbTi alloys\cite{nbti}, then one obtains a time
constant $\tau_0 \sim 2\times 10^{-4}$ s.  We now discuss the implications
of these estimates.

\section{Discussion}

If the above value of $\tau_0$ is substituted into our calculated frequency
spectrum, we find that the spectrum has a power-law frequency dependence
over a slightly different frequency range than 
that in which power-law behavior is seen
by Field {\it et al} in $Nb_{0.47}Ti_{0.53}$
(cf.\ Figs.~\ref{fig:pow} and~\ref{fig:knee}).  
Furthermore, the strength of the model flux noise power is smaller than
the experimental values by a factor ranging from 
two (at high frequencies, for high injection rate) to five (at low
frequencies, for low injection rate) orders of magnitude.

These differences between our calculated power spectrum and the measured
spectrum are not surprising.  The model is a only a rough 
approximation to a real material, since it uses an artificial kinetics
rather than a more realistic (and more computer-intensive) dynamics for the
calculation.  Furthermore, because of the finite size of the simulation
samples, they produce significantly less flux noise
than would be generated by a real material.  Nevertheless,
our model does give the qualitatively correct behavior: it leads to a 
power-law exponent which seems to approach the observed value for a 
sufficiently low vortex
injection rate and a sufficiently large lattice.

The noise spectra deviate from power-law behavior at low frequencies.
Specifically, they all show a weak peak at low frequencies, followed by a
further decrease with diminishing frequency and finally an increase at 
still lower frequencies.  We believe that this peak occurs near the frequency
of a characteristic ``lattice resonance.''   Qualitatively, this frequency is
the ratio of a characteristic length, i.\ e., the linear dimension of the
lattice, and a characteristic time, which is the time 
required for the lattice to ``reload'' to its Bean critical state
between avalanches.  For a lattice of size $L_x = 64$, having
pinning parameters $(0.1,5.0,0.1)$ and an injection rate of one vortex per 
time step, the resonant frequency occurs at about 3Hz (using our estimates
for lattice constant and time step).  For the same parameters but
slower injection rate (one vortex per 100$\tau_0$), the noise spectrum still
shows a peak but now at a lower resonance frequency of 0.3 Hz.  
For other values of the pinning parameters and injection rates, the resonance
becomes less conspicuous.  The resonant peak also becomes much less conspicuous
in a larger lattice.  In this case, the lattice simply contains more vortices.
Since some of these may not join the primary avalanche, they tend to
move separately, reducing the apparent reload time and hence producing
a stronger low-frequency noise signal.  The same argument holds 
when the pinning strength is increased at constant lattice size.  In this
case, there are once again more vortices in the system than at weaker pinning,
and hence, a stronger low-frequency flux noise signal.
Finally, at high injection rates (e. g., in the L$_x$ = 32 lattice at an
injection rate of one vortex per time step), the vortex system is in a state
of continual avalanche motion, since this rate is very large for such a small
lattice.  As a result, the power spectrum, which reflects that of the
avalanches, is relatively flat as a function of frequency.

Finally, we comment briefly on the relation of our calculation to a
similar study done by Nori and collaborators\cite{NoriC}.  In contrast
to our work, their calculations were carried out using dynamical equations
for individual vortices (assuming overdamped motion and a particular force
law to describe vortex-vortex interactions; the equations were solved using
molecular dynamics methods.  Their calculations were carried out solely
in the slow-driving limit; the resulting noise spectra, like those presented
here in the slow-driving limit, exhibited a power law in agreement with
experiments carried out in that slow-driving regime.
But our calculations do have the advantage of
speed: a realistic MD model using overdamped dynamics of individual
vortices requires approximately $10^4$ hours on a machine with
parallel processing.  By contrast, our
power spectra were calculated 
using a single Digital Alphastation 255.
for only $10^2$ hours.  This speed allows us to vary the parameters
extensively, in particular examining the effects of different driving
rates, which in turn permits investigation of the shoulders in the
power spectra mentioned above.  Of course, the kinetic approach does have
the compensating disadvantage of invoking a drastic simplification to
the true equations of motion; but it still appears to preserve much of
the relevant physics.

To summarize, we have shown that a relatively simple kinetic model
of vortex avalanches near the Bean critical state\cite{BP} gives rise
to flux noise which qualitatively resembles experiment, without using
computer-intensive dynamical equations for individual vortices.  Specifically,
the model predicts a power law flux noise spectrum with approximately the
correct power-law exponents, as well as a length-scale-induced knee
in the power spectrum for a sufficiently small lattice at high frequencies.
Although the model calculations differ in detail from experimental
results, they do show many qualitative similarities and trends in the
context of rather simple kinetic equations.

\section{Acknowledgments}

We are grateful for support from the National Science Foundation, Grant
DMR97-31511, and from the Department of Energy through the Midwest
Superconductivity Consortium at Purdue University, Grant No. DE-FG 02-90
ER45427.  We also thank F. Nori for valuable conversations.

\newpage

\begin{center}
{\bf Table I: Calculated lattice constant $a$ and magnetization density
$m(0)$ for several lattice sizes and two choices of model parameters}
\end{center}

\vspace{0.1in}

\begin{tabular}{||l|l|l|l||}
\hline
L& $m(0) (\Phi_0)$&$a (10^{-5}cm)$& slope$^\dagger$\\	\hline
$128^a$& 7500& 3.22& 1.83 \\ \hline
$64^a$& 1900& 3.24& 1.86 \\ \hline
$64^b$& 2300& (3.45)& 2.25 \\ \hline
$32^a$& 510& 3.37& 1.99 \\ \hline
\end{tabular}

\vspace{0.1in}

\noindent
$^a$parameters (r, p, q) = (0.1, 5.0, 0.1) as defined in text; \\
$^b$parameters (0.1, 12.0, 0.1).  The lattice constant for $b$ was
calculated using the same critical current as for $a$.  In actuality,
the larger pinning strength would call for a larger critical current.\\
$^\dagger$the slope is calculated from the columnar vortex density versus
distance into the sample, in terms of unit cells; \it{i.e.}
$(m(L_x)/L_y-m(0)/L_y)/(L_x-0)$. \\

\newpage

\begin{figure}
	\caption{Hexagonal lattice, with a rectangular overlay
displaying the unit cells used in the calculations.  Our ``unit cell''
is a rectangle enclosing four lattice points as shown.}
	\label{fig:lattice}
\end{figure}

\begin{figure}
        \caption{Cross-section of the vortex population in the sample
(schematically indicated by the length of the vertical arrows) after the vortex 
density has relaxed to the steady-state vortex profile, 
showing the linear decrease in magnetic induction with distance into the
sample, as predicted by the Bean model.  The path of the line integral
used to integrate Ampere's Law
is shown in boldface.}

	\label{fig:bean}
\end{figure}

\begin{figure}
	\caption{Schematic of experimental arrangement for noise measurements,
as viewed from above.  The arrows denote the average direction of vortex
motion, under the influence of an increasing external magnetic field applied
to the outside of the torus.  The coil is used to measure the flux noise.}
	\label{fig:sample}
\end{figure}

\begin{figure}
	\caption{Calculated voltage noise spectra due to flux motion into the
interior of the torus, for several choices of parameters and primarily for
larger lattices.  $r$ and $q$ are 0.1 for all spectra.  The legend
indicates the pinning strength,
$p$, and the lattice width (in units of $a_x$.)  The injection 
rate is one vortex injected per time step (fast injection) for
spectra $a$, $b$, and $c$,
and one vortex injected per 100 time steps (slow injection)for spectrum $d$.
The numerical data were collected for
a run of $2^{20}$ time steps, using a Savitzky-Golay smoothing filter
of order 1, using 40 points.  The translation into real frequency is
made assuming the estimates of time constant and lattice constant
as described in the text.}
	\label{fig:pow}
\end{figure} 

\begin{figure}
	\caption{Additional calculated vortex noise spectra, primarily for
smaller lattices.  These results show a possible ``knee'' in curve $a$.
This curve represents
an injection rate of one vortex per time step, which would correspond to 
experimental data taken at a high rate of vortex injection.
This curve should be contrasted to curves $b$ and $c$, which are taken at
lower rates of magnetic field ramping (1 vortex per 100 times steps) and show no
 such knee.}

	\label{fig:knee}
\end{figure}

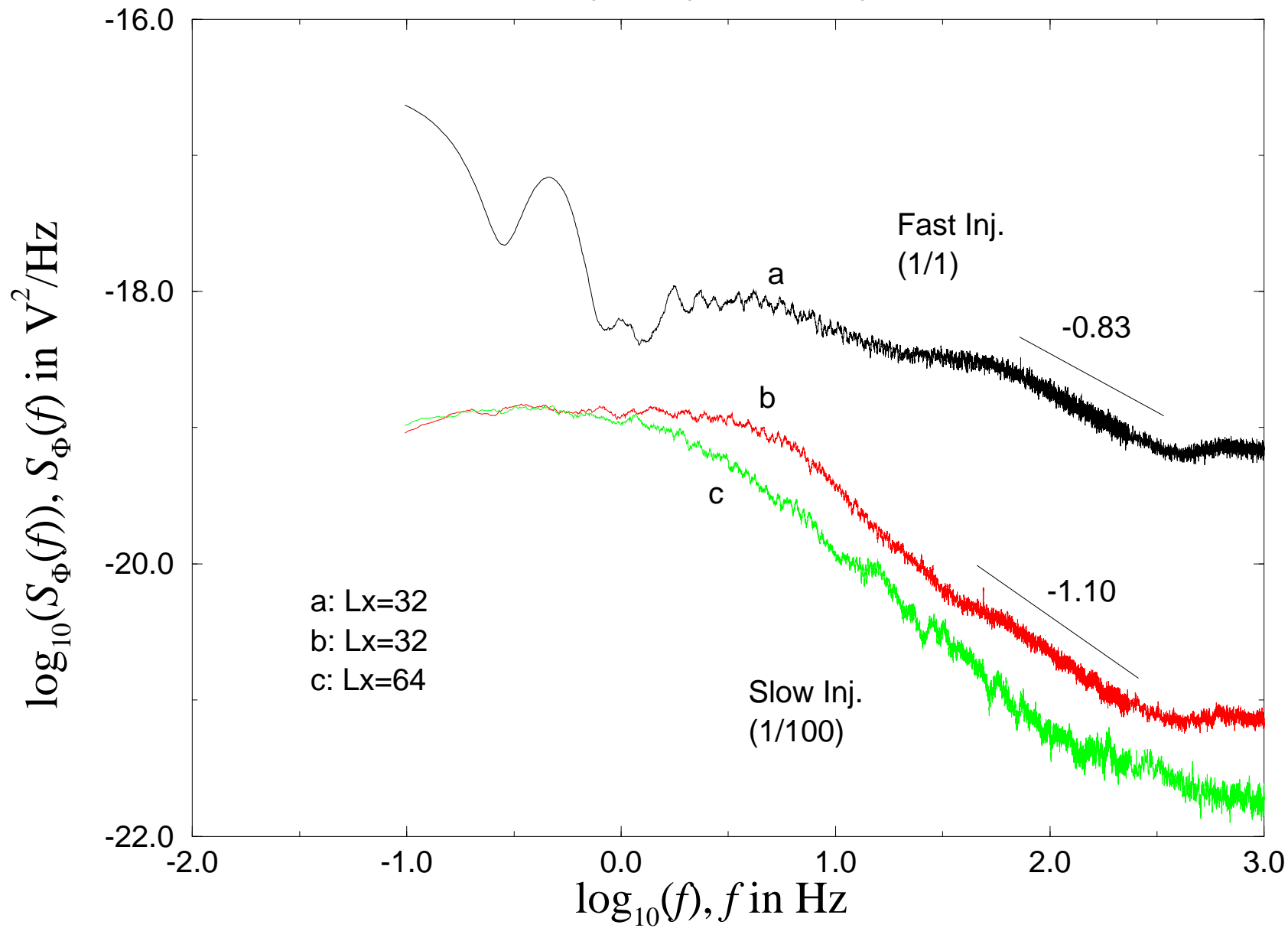
\end{document}